# PESSIMISM AND CYNICISM AS PHYSICAL ENTITIES

Victor Peterson II[1]

## ABSTRACT

Choosing pessimism makes one cynical to and necessitates their destruction of information.

A debate between pessimism and optimism has taken hold and structured the analytic landscape of cultural studies. Below, we treat this debate as a physical system. There's always been the question of how to formalize an account for the physical, material, connection between a model and the cultural phenomenon it explains. Lately, the concept of pessimism has garnered more attention than optimism. A connection between information and knowledge allows us to treat pessimism as a physical entity through the mathematical theory of information. Information theory utilizes the concept of entropy to discuss the reduction of alternatives with each signal received, allowing us to treat this transfer in physical terms. We consider the capacity to convert information to knowledge by way of an encoding/decoding process. (Hall 1973) A relation between features of phenomenal experience is encoded, indexing a concept. That frame is projected into subsequent contexts, organizing future experience so that the output received from using that frame can be decoded (=explained) with respect to that previously encoded concept. Analog experience, amplitudes of sensation registering over thresholds calibrated by past experience, is converted into decidable (=digital) categories (=encoded concepts). These categories are continually updated as prior cognitive output organizes what is of interest later on. Deciding on a category forgoes extraneous information. What we gain from that loss of information is the capacity to manipulate our environment. This capacity, then, is a part of the world although not a thing in it, for no object can be raised from the world to deny this capacity as it is the way we gain access to the world, fashioning world-views from that material. Our primary concepts adapt and are associated and combined to form more complex systems as our cognitive output organizes what future phenomenal input comes available relative to output from prior actions.

The concept of knowledge, then, is information produced belief. (Dretske 1981) The information travelling over some channel is encoded as being one thing despite that signal carrying other information. Only some features of that signal encodes a concept for the receiver. However, concepts can be applied in a way that one can generate false beliefs, despite, with respect to the encoding/decoding mechanism used, that concept being correct for that subject given the conditions to which it was indexed by that subject. A subject can believe that a dark room is empty while peering through a keyhole, for black is encoded as empty, but doesn't know that it is empty although the information traveling through that keyhole carries with it the information that there are entities there in the deep. There has been a coding error. The information that sustains that belief, that black is empty, is a subset of the information that lack of light may also mean that one cannot see the busy scene in the shadows. How, then, can the receiver *know* with *certainty* that the object of their belief is actually the case?

---

[1] RACE.ED Fellow at the Institute for Advanced Studies in the Humanities.

Information theory models the reduction of possibilities of a set of options over a channel. Given 10 options, each yes/no decision reduces alternatives. Ten options carry with it +3bits of information as it takes 3 rounds of questions to reduce alternatives to approximately 1. The link between probability and information means that this connection models that of entropy, licensing the treatment of information in physical terms. Therefore,

[1] $\quad I(X) = \sum_{i-1}^{n} p(x_i) \log_2 p(x_i)$

where $p$ is the probability of some $x_i$ of $X$ occurring coupled with the rate of reducing uncertainty with each yes/no determination within the domain (hence $\log_2$) until that quantity is reduced to a solution. The first term, the instance; the second, the information it carries. Flipping a coin encodes 1bit of information as there are 2 options and with one flip a decision is made, heads or tails, each previously having equal probabilities.

It would be a mistake to conflate information with meaning. The same concept, as a function of its application, can express different meanings given the different contexts following the conditions indexed by that function in which its use comes available. Meanings can evolve, whereas information is objective. In Zora Neale Hurston's *Characteristics of Expression*, information's currency. Encoding, projection, and decoding does not require interpretation although, apart from meaningful investment, would be informationally void. Meaning is made, information is its raw material. The connection between information (=input) and knowledge (=output) via the encoding of information is a concept that when projected into a domain organizes what input is available for the de/re-encoding of that domain in terms relevant to the receiver, can be made in the following way:

[2]     A subject $S$ knows $a$ is $F$ if and only if $S$'s belief that $a$ is $F$ is caused (sustained) by the information that $a$ is $F$.

The concept, "$x$ is $F$," encodes information indexing that concept to the conditions in which $F$-propositions apply. What *is* $F$ depends on the relation that object obtains in the context in which it appears, satisfying $F$-conditions although possibly appearing differently from what $F$ was abstracted from originally. The object of that proposition, then, is its function of its application. Functions defined as:

[3]   A function $f$ objectively determines a relationship where $x$ indexes a set of conditions and $y$ (=$x$-successor) indexes a context of application, such that the pair $(x,y)$ are members of $f$. When $x$ indexes conditions and the pair $(x,y)$ and $(x,y\text{-successor})$ are both members of $f$, then that $y$ and $y$-successor are functionally-equivalent.

The function $f$ becomes an abstract object. (Taylor 1998) It indexes a concept's conditions of application where $f(x)=x$ encodes $X$ at its zero, prior to use, and determines what counts in an $X$-domain, which is to say $X(0)=X$. Some $X$-proposition, an $X$-successor=$X_1$, references the conditions of its application through a line of citation up to $X_n$. If $Y$ represents a context of assertion, and if $X_1 \neq X$, therefore $X_1=Y$, then when $Y$ of $X$, $X$ is the name of the concept and $Y$ the context of application. As such, $Y(X)$ is appropriate. (Fara 2015) Functional-equivalence explains how the same function obtains of different things across contexts. Indirectly, we find that the same thing can obtain different functions.

The information that $X$ is the case is calculated by the average contribution of each $x_i$ contributing to the amount of information $X$ carries, i.e. the number of decisions to make until we arrive at *which $x$*. But how do we know, with certainty, which $x$? Can we be 100% certain? Apparently, uncertainty seems foundational for knowledge. If a concept is applicable universally, it contributes no new information, for we *know* not where and when it applies. An alternative with 0% possibility is undefined, cannot produce knowledge. Claude Shannon in developing information theory would say that absolutism leads to cynicism, for certainty annihilates information. How so?

We alluded to pessimism being treated as a physical entity by considering it a channel through which information flows. Gleaned from my conversations with the poet Donovan Munro, etymologically speaking, pessimism is marked as the greatest point of corrosion within that channel. For pessimists, the investiture of significance by the receiver fails, causing a collapse of certainty, (Gilroy 2004: 6) inducing a cynicism towards what's there ultimately leading to nihilism. How does the pessimist know that its object of interest is empty?

Consider the case between optimism (= alternatives must exist) and pessimism (= alternatives must not exist). Interestingly, both positions attribute 100% certainty to the existence of the alternatives within a state of affairs. This leads to an informational paradox for no new information is generated by an event with no alternatives. By [1]

[4]   An $S$-expression ($s_i$) such that the probability $p$ of $s_i$ approaches unity with the state of affairs itself, i.e. approaches 1, makes $\log p(s_i)$ go to 0.

If a state is necessarily determined one way, then regardless of how often that state's conditions are produced, if coupled with the decision that it can be no other way, then no information is associated with that state. As information is the condition sustaining what we know, and it is by what we know that we invest a state with meaning, the annihilation of knowledge by virtue of absolute certainty entails that no meaning is possible. The road to nihilism from cynicism passes through pessimism.

This information paradox has been explored in physical terms before. (Hawking 1975) If we suppose that nothing can travel faster than light, then a signal travelling at that speed or less generates no new information. If the volume of the universe can be calculated as the integral of the circumference of a sphere surrounding entities at various distances from a central point, then a black hole is a sphere within that volume that has infinite radius. As such, black holes seemingly destroy information, not even light escapes. However, if a photon falls into that hole, a hole that's by definition a vacuum, the energy it carries should wholly transfer into the state circumscribed by that sphere. Yet, as black holes are defined as having a negative infinite radius, the fall to the surface never fully occurs, leaving a trace. (S. Haco, et al 2108) This is illuminating for the cynic must maintain (=record) that which allows it to *know* that the state is the way they've determined in order to maintain their own position. The cynic cannot prove their cynicism as they would have to produce an entity from that domain to show it's empty. However, the function of expressing some entity, if identical to that entity, means that its function of expression is an object in itself, showing that that domain is not empty after all.

Another example is that of interest convergence. (Bell Jr. 1980) This theory states that convergent interests lead to the maintenance of the channel, not the information, thereby reinforcing what's understood as the norm or dominant position. Cynicism is understood by absolutism of frame. A channel, by definition, is the sum alternatives that don't contribute information, setting the boundaries through which information flows, i.e. what is fixed relative to what varies thereby contextualizing what can be known. Convergence converts everything that flows through a channel to the same information. Independent alternatives are combined to fit either a category that maintains the flow of information in the same direction or to a category that is fed back into the system.

Compositionally, if $X$ is the category to be returned and $Y$ to be retained, with $I(Z)$ contributing to $X$, and $f(Z) = \prod_{i-1}^{n} p(z_1 \ldots z_n)^{Np(x_n)} = \sum_{i-1}^{n} p(x_i) log p(z_i)$, with what's $X$ expressed as a function of $N$ occurrences, then

[1] If $X=f(Z)$, then $I(f(Z)$ given $Z)=0$, and $I(Z, f(Z))$ evaluates $X$ to $I(Z)+I(f(Z)$ given $Z)=I(Z)+I(Z$ given $f(Z))$, meaning that $I(f(Z)) \leq I(Z)$, diminishing the content of $X$ to null before it produces output.

With $f$ being recursive, $Z$ diminishes towards 0 as it's returned as input after passing through the channel and multiplied over $X^N$ many iterations. The output of the channel is maintained insofar as

[2] $\quad C(X,Y) = I(Y) + I(X) = \sum_y p(y) log p(y) + \sum_{i-1}^{n} p(x_i) log p(z_i)$

where to the right of the addition sign, the first term of the product sum represents the probability of categorization and the second the information produced/predicated. $X$ is received although $Z$ was sent. Alternatives arise but are ignored. If accepted, the information contributes (=additive); if not, it's recursively coupled with the unacceptable (=multiplied). The total sum of the combination of probable alternatives reduces to 0 after a certain number of cycles. The channel annihilates information that doesn't contribute to the current structure of the state, deeming certain sources as nonexistent regardless of what's transmitted. This nihilistic outcome is derived from the need to evacuate a source of content upon arrival so as to maintain the current state. However, if 0 is received, this doesn't mean that nothing was sent, in fact, we find that that is due to the channel, not the source.

This misinterpretation grounds Afropessimism, a term originating in International Relations that described the futility of investments in Africa. (Hawkins 1990; Hitchens 1994) It's now conflated with the impossibility of Blacks contributing to global socio-political affairs. (Wilderson III 2020) However, we do find evidence of culturally and politically significant transmission despite its negation upon arrival. (Mineo L. and Patterson O. 2018) We see that it's the historical formation and current structure of the channel—slavery, colonization, appropriation—and not the source that is the blame. As a 0% possibility is undefined information-theoretically, it is not necessarily the case that Black-ness has no import, only that it's necessarily barred from that channel and, possibly, is integral to it if it is the case that that probability does not vary given changing conditions. With 0% possibility being undefined information-theoretically, the limits of this archaic channel indirectly proves the necessary possibility of alternative channels.

Reduction to statistical combinations of behavior as a proof of their cause denies the historical formation of structures projected to organize, allowing or disallowing, certain expressions as a function of prior output. The pessimist channel decides what doesn't contribute information, only attributing existence to what passes. Cynicism emerges. We've found that pessimism cannot *know* that its object is void. Their disposition would be that of a cynic, then, not a pessimist. We find that nothing is certain, even the certainty that everything means nothing. Channel fundamentalism results from the cynicism of the pessimist deciding nothing is to be gained from a vested interest in the very subject they analyze. Information-theoretically, the cynicism of the pessimist, and its resulting nihilism are not default positions—see Gilroy; they emerge given the channel through which information is transmitted. Attend to the channel, alternatives arise.